\documentclass[useAMS,usenatbib,referee]{biom}

\usepackage{verbatim}
\usepackage{natbib}
\usepackage{url} % not crucial - just used below for the URL 
\usepackage{amsmath}
\usepackage{natbib}
\usepackage[colorlinks=true,allcolors=blue]{hyperref}
\usepackage{url}
\usepackage{graphicx}
\usepackage{amsmath, amssymb}
\usepackage{mathrsfs} 
\usepackage{bbold}
\usepackage{mathbbol}
\usepackage{float}
\usepackage{xcolor}
\usepackage{bm} % Math packages
\usepackage{setspace}
\usepackage{multirow}
\usepackage{lscape}
\usepackage[figuresright]{rotating}
\usepackage{adjustbox}
\usepackage{booktabs}
\usepackage{ragged2e}
\usepackage{algorithm}
\usepackage{algorithmic}
\usepackage{caption}
\usepackage{subcaption}
\usepackage{tikz}
\usepackage{float}
\def\*#1{\mathbf{#1}}
\def\+#1{\amsmathbb{#1}}
\def\##1{\mathbb{#1}}
\DeclareSymbolFontAlphabet{\amsmathbb}{AMSb}%

\def\bSig\mathbf{\Sigma}

\title[]{Variable Selection in Functional Linear Cox Model}

\author
{Yuanzhen Yue$^{1}$, Stella Self$^{1}$, Yichao Wu$^{2}$, Jiajia Zhang$^{1}$, and Rahul Ghosal$^{1,*}\email{rghosal@mailbox.sc.edu}$ \\
$^{1}$Department of Epidemiology and Biostatistics, University of South Carolina. \\
$^{2}$Department of Mathematics, Statistics, and Computer Science, University of Illinois at Chicago.}

\begin{document}

%  This will produce the submission and review information that appears
%  right after the reference section.  Of course, it will be unknown when
%  you submit your article, so you can either leave this out or put in 
%  sample dates (these will have no effect on the fate of your article in the
%  review process!)

%\date{{\it Received October} 2007. {\it Revised February} 2008.  {\it
%Accepted March} 2008.}

%  These options will count the number of pages and provide volume
%  and date information in the upper left hand corner of the top of the 
%  first page as in published articles.  The \pagerange command will only
%  work if you place the command \label{firstpage} near the beginning
%  of the document and \label{lastpage} at the end of the document, as we
%  have done in this template.

%  Again, putting a volume number and date is for your own amusement and
%  has no bearing on what actually happens to your article!  

%\pagerange{\pageref{firstpage}--\pageref{lastpage}} 
%\volume{64}
%\pubyear{2008}
%\artmonth{December}

%  The \doi command is where the DOI for your article would be placed should it
%  be published.  Again, if you make one up and stick it here, it means 
%  nothing!

%\doi{10.1111/j.1541-0420.2005.00454.x}

%  This label and the label ``lastpage'' are used by the \pagerange
%  command above to give the page range for the article.  You may have 
%  to process the document twice to get this to match up with what you 
%  expect.  When using the referee option, this will not count the pages
%  with tables and figures.  

\label{firstpage}

%  put the summary for your article here

\begin{abstract}
% As biomedical studies increasingly gather complex, high-dimensional physiological data, effective variable selection methods are essential to manage this complexity and enhance accuracy in survival models.
% Additionally, we introduce a novel framework for selecting smoothing parameters within the Extended Bayesian Information Criteria (EBIC), distinguished by a new method for calculating degrees of freedom.
Modern biomedical studies frequently collect complex, high-dimensional physiological signals using wearables and sensors along with time-to-event outcomes, making efficient variable selection methods crucial for interpretation and improving the accuracy of survival models. We propose a novel variable selection method for a functional linear Cox model with multiple functional and scalar covariates measured at baseline. We utilize a spline-based semiparametric estimation approach for the functional coefficients and a group minimax concave type penalty (MCP), which effectively integrates smoothness and sparsity into the estimation of functional coefficients. An efficient group descent algorithm is used for optimization, and an automated procedure is provided to select optimal values of the smoothing and sparsity parameters. Through simulation studies, we demonstrate the method’s ability to perform accurate variable selection and estimation. The method is applied to 2003-06 cohort of the National Health and Nutrition Examination Survey (NHANES) data, identifying the key temporally varying distributional patterns of physical activity and demographic predictors related to all-cause mortality. Our analysis sheds light on the intricate association between daily distributional patterns of physical activity and all-cause mortality among older US adults.  \\
\end{abstract}

%  Please place your key words in alphabetical order, separated
%  by semicolons, with the first letter of the first word capitalized,
%  and a period at the end of the list.
%


%  As usual, the \maketitle command creates the title and author/affiliations
%  display 

\maketitle

%  If you are using the referee option, a new page, numbered page 1, will
%  start after the summary and keywords.  The page numbers thus count the
%  number of pages of your manuscript in the preferred submission style.
%  Remember, ``Normally, regular articles exceeding 25 pages and Reader Reaction 
%  articles exceeding 12 pages in (the preferred style) will be returned to 
%  the authors without review. The page limit includes acknowledgements, 
%  references, and appendices, but not tables and figures. The page count does 
%  not include the title page and abstract. A maximum of six (6) tables or 
%  figures combined is often required.''

%  You may now place the substance of your manuscript here.  Please use
%  the \section, \subsection, etc commands as described in the user guide.
%  Please use \label and \ref commands to cross-reference sections, equations,
%  tables, figures, etc.
%
%  Please DO NOT attempt to reformat the style of equation numbering!
%  For that matter, please do not attempt to redefine anything!

\section{Introduction}
\label{sec:intro4}
Functional data analysis (FDA) \citep{Ramsay05functionaldata,crainiceanu2024functional} is a statistical framework used to analyze data that can be represented as functions, curves, or trajectories over a continuum such as time, space, or other domains. Unlike traditional multivariate analysis, which focuses on finite-dimensional vectors, FDA deals with infinite-dimensional objects, making it particularly suitable for complex datasets where observations are best understood as smooth curves or surfaces. The applications of FDA are vast and span numerous scientific fields. In the biomedical sciences, FDA is utilized to analyze growth curves \citep{ramsay2007applied}, heart rate \citep{ratcliffe2002functional,diller2006heart}, physical activity \citep{xiao2015quantifying,goldsmith2016new,cui2021additive,cui2022fast,ghosal2023functional}, and brain activity patterns \citep{tian2010functional} over time, offering deeper insights into physiological processes and health outcomes. Environmental studies benefit from FDA through the analysis of temporal and spatial data, such as climate patterns and pollution levels, allowing for the modeling and prediction of environmental phenomena \citep{besse2000autoregressive}. In finance, FDA is used to model and forecast economic indicators, stock prices, and interest rates, where the continuous nature of the data provides a richer understanding of market behaviors \citep{horvath2012inference}.

The integration of FDA and survival analysis is particularly valuable in biomedical research, where continuous monitoring of physiological signals is common. 
%Examples of such signals include heart rate \citep{diller2006heart}, electrocardiograms (ECG) \citep{bossone2002prognostic}, and physical activity \citep{cui2021additive,cui2022fast,ghosal2023functional} recorded over time. 
The functional measurements capture detailed temporal variations that may hold important prognostic value for survival outcomes, such as time to disease progression, recurrence, or death. By leveraging FDA, we can effectively model and analyze these functional covariates, enabling a more comprehensive understanding of their relationships with survival outcomes. Functional Cox models \citep{gellar2015cox,cui2021additive} extend the traditional Cox proportional hazards model \citep{cox1972regression} by incorporating functional covariates as predictors. Several approaches have been proposed for estimation and inference in the functional Cox model \citep{kong2018flcrm,hao2021semiparametric} and its extensions.

With advancements in technology, the complexity of biomedical data has increased dramatically, especially in modern studies involving high-dimensional physiological signals and time-to-event outcomes. These studies now routinely collect vast amounts of data, posing significant challenges in identifying the high-dimensional variables truly associated with the outcome of interest. Robust variable selection methods are therefore crucial for improving 
interpretation and accuracy of such high-dimensional models \citep{hastie2015statistical}. 
%\textcolor{red}{First survival, then scalar on function}
Several methods have been developed for variable selection in ``classical" survival models with scalar covariates based on penalized likelihood approaches such as least absolute shrinkage and selection operator (LASSO) \citep{tibshirani1997lasso,simon2011regularization}, elastic net \citep{wu2012elastic}, and other nonconcave extensions \citep{fan2002variable, du2010penalized,honda2014variable}. In scalar-on-function regression \citep{reiss2017methods,crainiceanu2024functional}, several methods have been proposed to perform variable selection that select the relevant functional predictors \citep{gertheiss2013variable,ma2016estimation,collazos2016consistent}, as well as estimate their smooth dynamic association with the scalar outcome of interest.

\begin{figure}[h!]
  \centering
    \begin{subfigure}{\textwidth}
        \centering
        \includegraphics[scale=0.47]{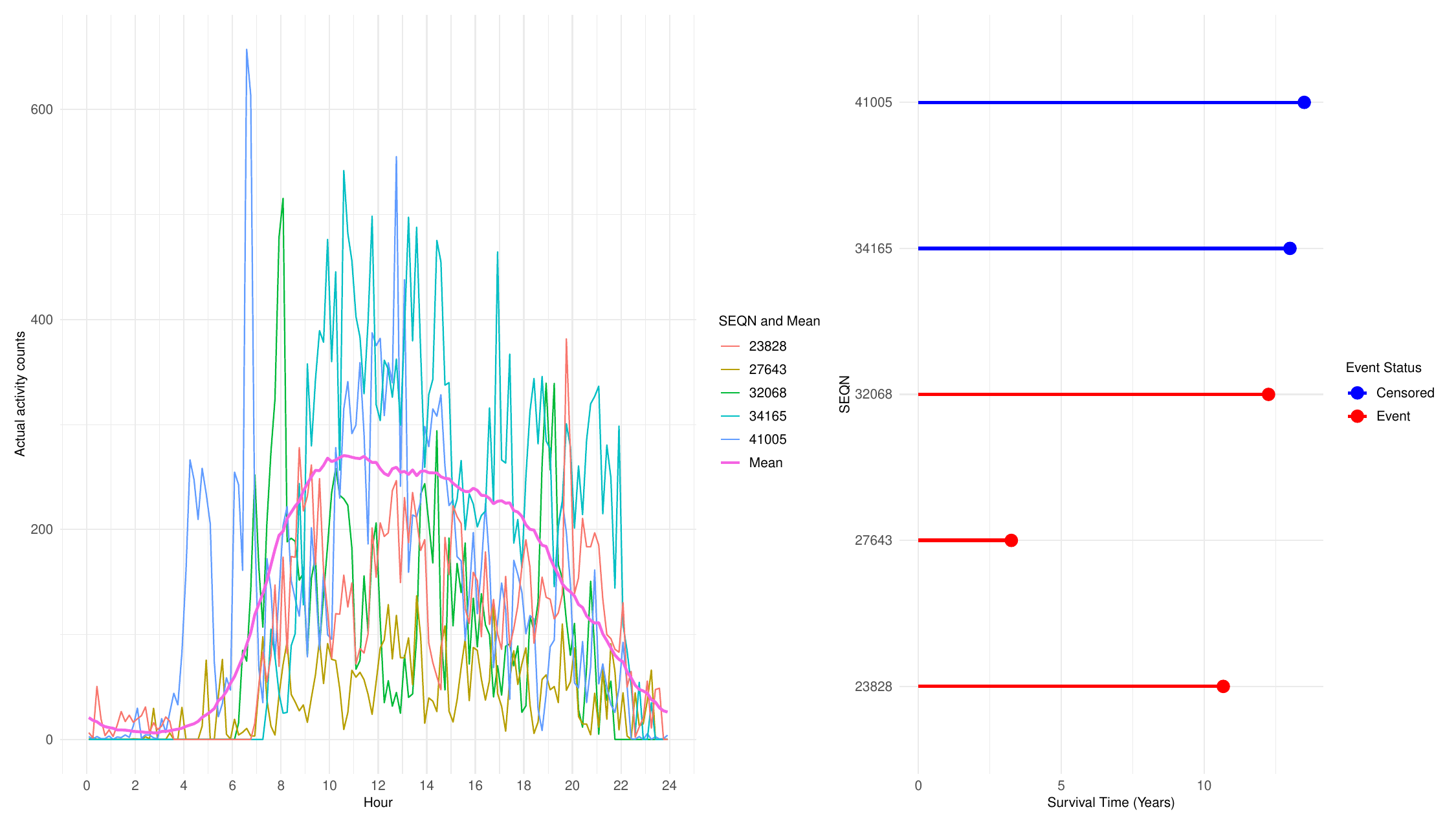}
        \caption{The average activity counts across all participants, along with the actual activity counts and survival time for five randomly selected participants by SEQN (unique subject identifier)}
        \label{1a}
    \end{subfigure}
    
    \begin{subfigure}{\textwidth}
        \centering
        \includegraphics[scale=0.45]{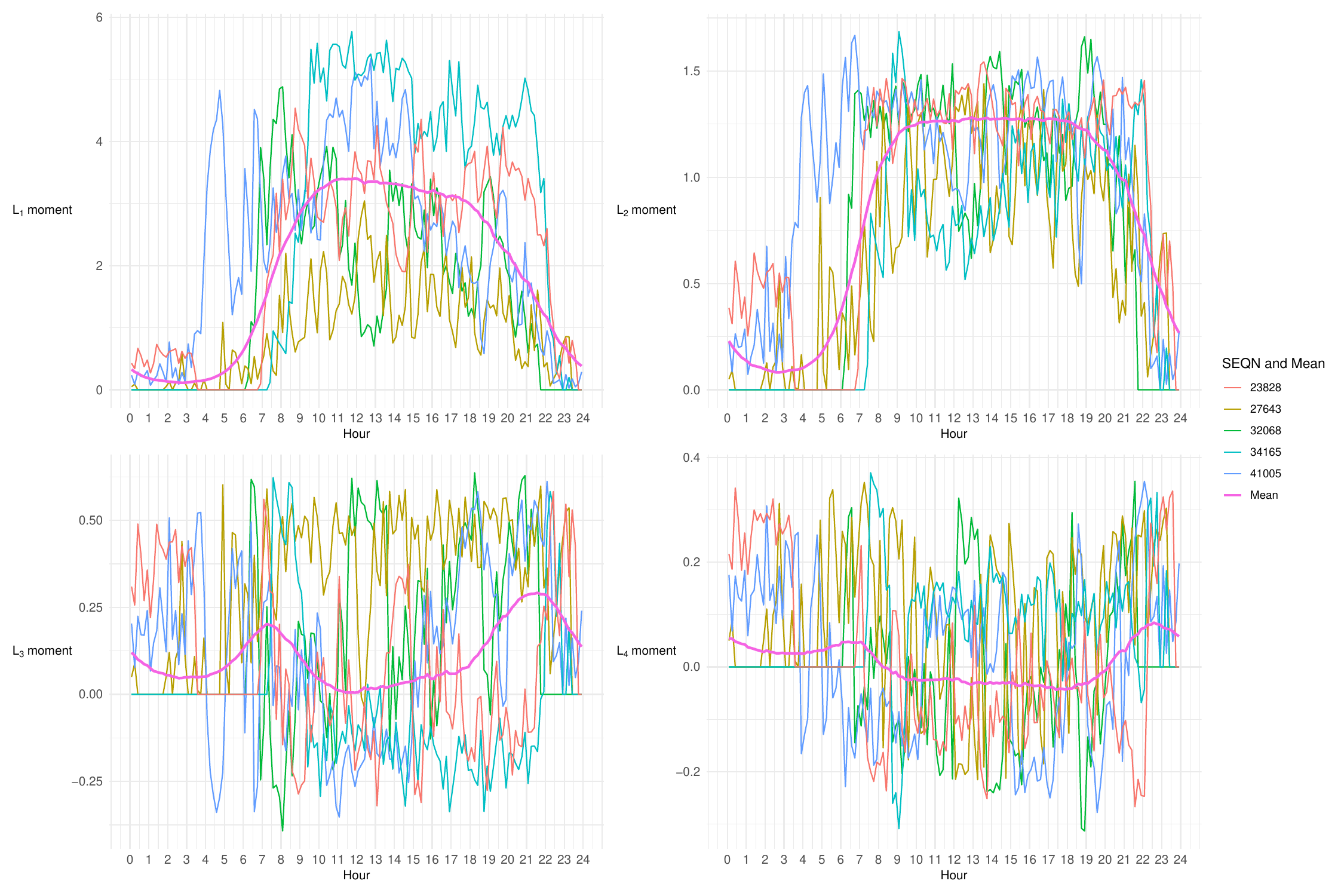}
        \caption{The average of the first four L-moments of the time-varying physical activity profiles across all participants, as well as the first four L-moments for five randomly selected participants by SEQN, using log-transformation}
        \label{1b}
    \end{subfigure}
\caption{Visual summary of physical activity across participants}
\label{fig1}
\end{figure}

Despite the increasing collection of high-dimensional functional observations, such as physical activity, heart rate, and energy expenditure, existing literature on variable selection in survival models that address the simultaneous selection of functional and scalar covariates is relatively sparse. In our motivating application, we specifically focus on accelerometer data from the 2003-06 waves of the National Health and Nutrition Examination Survey (NHANES). 
This dataset is linked with the National Death Index (NDI) up to December 31, 2019 \citep{leroux2019organizing} to define our survival outcome of interest. Numerous studies in the past in NHANES have demonstrated a consistent link between higher levels of physical activity (PA) and reduced mortality risk \citep{koster2012association,saint2020association}.
Traditionally, these models have been based on summary-level PA metrics \citep{leroux2019organizing,smirnova2020predictive, ledbetter2022cardiovascular,leroux2024nhanes}. Recent methods using a functional Cox model and its extensions have investigated associations of the average daily patterns \citep{xiao2015quantifying} of physical activity with mortality risk \citep{cui2021additive,ghosal2023functional}. Figure \ref{1a} displays the average diurnal physical activity counts across all participants, accompanied by detailed activity profiles for five randomly chosen individuals, along with their respective survival times and censoring statuses. 
Recent research in the area of distributional analysis \citep{ghosal2023distributional} has highlighted that distributional features beyond the mean, such as variability and other higher order moments of PA varying through the day, can provide meaningful and complementary information that can enhance insights beyond the average diurnal PA pattern \citep{varma2021continuous,ghosal2022scalar}. The daily time-of-day dependent L-moments \citep{ghosal2022scalar, cho2024exploring} provide a useful framework for characterizing the daily distributional patterns of physical activity. Figure \ref{1b} illustrates the average of the first four diurnal L-moments of physical activity profiles across all participants, as well as the first four diurnal L-moments for five randomly selected participants, based on the log-transformed activity counts. The main objective of our study is to identify the influential functional predictors from the diurnal higher order L-moments of PA, as well as the key demographic and lifestyle covariates which are associated with all-cause mortality among older US adults .

%scalar predictors among age, body mass index (BMI), and smoking status—as well as the , which are associated with all-cause mortality.
% \cite{cui2021additive} proposed the Additive Functional Cox Model, and \cite{ghosal2023functional} developed a functional proportional hazards mixture cure model. Their applications to NHANES data provided novel epidemiological insights into the association between daily patterns of physical activity and mortality.

In this paper, we develop a method for variable selection in a functional linear Cox model (FLCM)  with multiple functional and scalar covariates measured at the baseline. Our main contributions are three-fold. First, we demonstrate that in a functional linear Cox model with multiple functional and scalar covariates, the variable selection problem can be identified as a group selection problem.  Second, we utilize a spline-based semiparametric estimation approach for the functional coefficients and a group minimax concave type penalty
(MCP), which jointly enforces smoothness and sparsity into the estimation of functional coefficients, along with the selection of the scalar covariates. Third, we employ an efficient group descent algorithm for optimization and provide an automatic procedure for choosing the tuning parameters based on an extended Bayesian information type criterion (EBIC) \citep{chen2008extended}.
% We have also developed an innovative method for calculating degrees of freedom, used within the context of Extended Bayesian Information Criteria (EBIC) \citep{chen2008extended} to select both tuning and penalty parameters.

The rest of this paper is organized as follows. In Section \ref{sec:method1}, we present our modeling framework and illustrate our variable-selection method. In Section \ref{sec:sim_stud}, we conduct a simulation study to evaluate the performance of our method and summarize the results. In Section \ref{realdat}, we apply the proposed variable-selection method to the NHANES 2003-06 accelerometer data and present our findings. We conclude in Section \ref{disc} with a discussion and deliberate on some potential extensions of our work.
\vspace{- 9 mm}

\section{Methodology}
\label{sec:method1}
\subsection{Modeling Framework}
\label{mf1}
We denote by $T_i$ the survival time for subject $i$, and $C_i$ the corresponding censoring time for 
$i=1,\ldots,n$. The observed survival time is given by $Y_i= \min(T_i, C_i)$ in the presence of right censoring, and $\Delta_i= I(T_i\leq C_i)$ is the event indicator. The scalar baseline covariates are denoted as $ \boldsymbol{Z_{i}}=( Z_{i1}, Z_{i2}, \ldots, Z_{ip} )^T$. We also observe functional covariates, \(Z_{ik}(s) \in \mathscr{L}^2 [0, 1]\), for \(k = 1, 2, \ldots, K\), for each subject at the baseline. We assume they lie in a real separable Hilbert space, taken to be $\mathscr{L}^2 [0, 1]$ in this paper. In developing our method, we assume these functional covariates are observed on a dense and regular grid of points \(S = \{t_{1}, t_{2}, \ldots, t_{m}\} \subset [0, 1]\), matching our motivating application. Although, this can be relaxed and extended to more general scenarios.
Denote the data observed for the \(i\)th subject, as $D_i=\{T_i, C_i, Y_i, \delta_i, \boldsymbol{Z_{i}},Z_{ik}(s)\}$ (for $i=1,2,\ldots,n$). We assume that the survival time $T$ is independent of the censoring time $C$, conditional on the scalar and functional covariates.

\subsection{Functional Linear Cox Model}
We posit a functional linear cox model \citep{gellar2015cox} given by:
\begin{equation}\label{m}
\log h_i(t; \boldsymbol{\beta}, \boldsymbol{\beta}(\cdot)) = \log h_0(t) + \boldsymbol{Z_{i}}^T \boldsymbol{\beta} + \sum_{k=1}^{K} \int_0^1 Z_{ik}(s) \beta_k(s) \, ds,
\end{equation}
where \(h_i(t; \boldsymbol{\beta}, \boldsymbol{\beta}(\cdot))\) is the hazard at time \(t\) given covariates \(\boldsymbol{Z_{i}}\) and \(\{Z_{ik}(s), k = 1, 2, \ldots, K\}\), \(h_0(t)\) is the baseline hazard function, \(\boldsymbol{\beta} = (\beta_1, \beta_2, \cdots, \beta_p)^T\) are scalar parameters associated with log hazard ratio of the scalar covariates. The functional parameters \(\beta_k(s)\) (for \(k = 1, 2, \ldots, K\)) are assumed to be smooth with finite second derivatives, and capture the association with the log hazard at index $s$. Specifically, a positive value of \( \beta_k(s) \) indicates that an increase in the corresponding functional covariate at time \( s \) is associated with an increase in the log hazard, assuming all other covariates remain unchanged.

We model the unknown coefficient functions \( \beta_k(s) \) using a set of known basis functions \(\{\theta_{kc}(s), c = 1, 2, \ldots, C_k\}\) for \(k = 1, 2, \ldots, K\). The coefficient functions are thus expressed through a basis function expansion as:
\begin{equation*}
    \beta_k(s) = \sum_{c=1}^{C_k} b_{kc} \theta_{kc}(s) = \boldsymbol{\theta}_k(s)^T \boldsymbol{b}_k ,
\end{equation*}
where \(\boldsymbol{\theta}_k(s) = (\theta_{k1}(s), \theta_{k2}(s), \ldots, \theta_{k C_k}(s))^T\), and \(\boldsymbol{b}_k = (b_{k1}, b_{k2}, \ldots, b_{k C_k})^T\) is a vector of unknown coefficients. Throughout this paper, we employ B-spline basis functions, though other basis functions may also be used.

We use a penalized partial log-likelihood approach \citep{gellar2015cox,cui2021additive} for estimation of the unknown parameters. The partial likelihood \citep{gellar2015cox} for a functional linear Cox model (FLCM) is given by
\begin{equation}
L(\boldsymbol{\beta}, \boldsymbol{\beta}(\cdot)) = \prod_{d \in D} \frac{\exp(\eta_d)}{\sum_{r \in R_d} \exp(\eta_r)},
\end{equation}
where \(D\) is the set of failure indices, \(R_d\) is the set of individuals at risk at time \(t_d\), 
and $\eta_i$ is the systematic component of our model defined as: \(\eta_i = \sum_{j=1}^{p} Z_{ij}\beta_j + \sum_{k=1}^{K} \int_0^1 Z_{ik}(s) \beta_k(s) \, ds\). The log partial likelihood is denoted by: $\ell(\boldsymbol{\beta}, \boldsymbol{\beta}(\cdot)) = \log L(\boldsymbol{\beta}, \boldsymbol{\beta}(\cdot))$.

Since, we are interested in smooth estimation of the functional coefficients \citep{gellar2015cox} as well as imposing sparsity \citep{gertheiss2013variable,ghosal2020variable}, we impose a penalty on the log partial likelihood to ensure both sparsity and smoothness. 
%The application of penalized log partial likelihood in Cox models is well-established. For example, \cite{verweij1994penalized} and \cite{therneau2003penalized} utilized the established relationship between mixed effects models and penalized splines to estimate frailty models using the penalized log partial likelihood method.
% The application of a penalized partial likelihood function in Cox models is well-established. \cite{verweij1994penalized} and \cite{therneau2003penalized} utilized the established relationship between mixed effects models and penalized splines to estimate frailty models using the penalized partial likelihood method. We adopt a similar approach to incorporate functional predictors into the Cox model. The partial likelihood for our model is:
% \begin{equation}
% L(\boldsymbol{\beta}, \boldsymbol{\beta}(\cdot)) = \prod_{d \in D} \frac{\exp(\eta_d)}{\sum_{r \in R_d} \exp(\eta_r)},
% \end{equation}
% where \(D\) is the set of failure indices, \(R_d\) is the set of individuals at risk at time \(t_d\), 
% and $\eta_i$ is the systematic component of our model defined as: \(\eta_i = \sum_{j=1}^{p} Z_{ij}\beta_j + \sum_{k=1}^{K} \int_0^1 Z_{ik}(s) \beta_k(s) \, ds\). Then the log partial likelihood is denoted by: $\ell(\boldsymbol{\beta}, \boldsymbol{\beta}(\cdot)) = \log L(\boldsymbol{\beta}, \boldsymbol{\beta}(\cdot))$.
We apply separate penalties to the scalar and functional parameters to account for the smoothness of the functional parameters. For scalar parameters, penalties like LASSO \citep{tibshirani1996regression}, MCP \citep{zhang2010nearly}, etc., can be applied for performing variable selection. For example, the LASSO penalty is defined as $P_{\lambda}\{\beta_j\} = \lambda \|\beta_j\|_1$, which effectively shrinks certain coefficients to zero, thereby eliminating non-informative predictors from the model. Since the functional coefficients are assumed to be smooth, we impose a penalty on the roughness of the coefficient functions \(\beta_k(s)\) as follows:
\begin{equation}\label{eq:6}
\begin{split}
P_{\lambda,\psi}\{\beta_k (\cdot)\} &= \lambda \left\{ \int \beta_k(\cdot)^2 \, dt + \psi \int \beta_k''(\cdot)^2 \, dt \right\}^{1/2} \\
&= \lambda \left( \boldsymbol{b}_k^T \mathbb{R}_k \boldsymbol{b}_k + \psi \boldsymbol{b}_k^T \mathbb{Q}_k \boldsymbol{b}_k \right)^{1/2} \\
&= \lambda \left( \boldsymbol{b}_k^T \mathbb{K}_{\psi,k} \boldsymbol{b}_k \right)^{1/2},
\end{split}
\end{equation}
where \(\mathbb{K}_{\psi,k} = \mathbb{R}_k + \psi \mathbb{Q}_k\), \(\mathbb{R}_k = \int \boldsymbol{\theta}_k(s) \boldsymbol{\theta}_k(s)^T \, dt\), and \(\mathbb{Q}_k = \int \boldsymbol{\theta}_k''(s) \boldsymbol{\theta}_k''(s)^T \, dt\). The parameter \(\psi \geq 0\) determines the degree of penalization for the roughness.

Using the Cholesky decomposition \(\mathbb{K}_{\psi,k} = \mathbb{L}_{\psi,k} \mathbb{L}_{\psi,k}^T\) \citep{gertheiss2013variable,ghosal2020variable} and defining \(\boldsymbol{\gamma}_k = \mathbb{L}_{\psi,k}^T \boldsymbol{b}_k\), equation (\ref{eq:6}) can be reparameterized as: $P_{\lambda,\psi}\{\beta_k (\cdot)\}=\lambda(\boldsymbol{\gamma}_k^T \boldsymbol{\gamma}_k)^{1/2}=\lambda \|\boldsymbol{\gamma}_k)\|_2$. The systematic component can now be reformulated as,
\begin{equation}
\begin{split}
    \eta_i &= \sum_{j=1}^{p} Z_{ij} \beta_j + \sum_{k=1}^{K} \int_0^1 Z_{ik}(s) \beta_k(s) \, ds \\
    &= \sum_{j=1}^{p} Z_{ij} \beta_j + \sum_{k=1}^{K} \int_0^1 Z_{ik}(s) \left( \sum_{c=1}^{C_k} b_{kc} \theta_{kc}(s) \right) \, ds \quad\\
    &= \sum_{j=1}^{p} Z_{ij} \beta_j + \sum_{k=1}^{K} \sum_{c=1}^{C_k} \left( b_{kc} \cdot \frac{1}{m} \sum_{s=0}^{1} Z_{ik}(s) \theta_{kc}(s) \right) \\
     &= \sum_{j=1}^{p} Z_{ij} \beta_j + \sum_{k=1}^{K} \sum_{c=1}^{C_k} \left( b_{kc} \cdot Z_{ikc}^{*} \right) \\
\end{split}
\end{equation}

\begin{equation}
%\label{eq:7}
\begin{split}
     \eta_i &= \sum_{j=1}^{p} Z_{ij} \beta_j + \sum_{k=1}^{K} \boldsymbol{Z}_{ik}^{*}{}^T \cdot \boldsymbol{b}_k \notag\\
    &= \sum_{j=1}^{p} Z_{ij} \beta_j + \sum_{k=1}^{K} \tilde{\boldsymbol{Z}}_{ik}{}^T \cdot \boldsymbol{\gamma}_k,
\end{split}
\end{equation}
where $m$ is the number of grid points over the interval $[0,1]$, \(Z_{ikc}^{*} = \frac{1}{m} \sum_{s=0}^{1} Z_{ik}(s) \theta_{kc}(s)\), \(\boldsymbol{Z}_{ik}^{*} = (Z_{ik1}^{*}, Z_{ik2}^{*}, \ldots, Z_{ikC_k}^{*})^T\), \(\tilde{\boldsymbol{Z}}_{ik} = \boldsymbol{Z}_{ik}^{*} \cdot (\mathbb{L}_{\psi,k}^T)^{-1}\) and \(\boldsymbol{\gamma}=(\boldsymbol{\gamma}_1^T,\ldots,\boldsymbol{\gamma}_K^T)^T\).

We recognize the variable selection in our functional linear Cox model can be addressed as a group selection problem, where the grouping is determined by the covariates $(\tilde{\boldsymbol{Z}}_{ik}$ (corresponding to each functional predictor). For example, we can obtain estimates of \(\boldsymbol{\beta}\) and \(\boldsymbol{\gamma}\) by minimizing a penalized log partial likelihood criterion with group LASSO penalty \citep{yuan2006model} on the basis coefficients, and a LASSO penalty on the coefficients of the scalar covariates,
\begin{equation}
\begin{split}
    (\hat{\boldsymbol{\beta}}, \hat{\boldsymbol{\gamma}}) &= \underset{\boldsymbol{\beta},\boldsymbol{\gamma}}{\arg \min} \left\{ \frac{1}{n} (-2 \ell(\boldsymbol{\beta}, \boldsymbol{\beta}(\cdot)) + \sum_{j=1}^{p} \lambda \|\beta_j\|_1 + \sum_{k=1}^{K} \lambda \|\boldsymbol{\gamma}_k\|_2) \right\} \\
    &= \underset{\boldsymbol{\beta},\boldsymbol{\gamma}}{\arg \min} \left\{ \frac{1}{n} (-2 \ell(\boldsymbol{\beta}, \boldsymbol{\beta}(\cdot)) + \sum_{j=1}^{p} P_{\text{LASSO}, \lambda} (\|\beta_j\|_1) + \sum_{k=1}^{K} P_{\text{LASSO}, \lambda} (\|\boldsymbol{\gamma}_k\|_2)) \right\}.
\end{split}
\end{equation}

\subsection{VSFCOX Method for Variable Selection in FLCM}

In this paper, we propose the VSFCOX method for performing variable selection in FLCM, by using a group minimax concave penalty (MCP) \citep{zhang2010nearly} extension of the above optimization criterion (5).  Although LASSO \citep{tibshirani1996regression} is a widely used penalization technique for high-dimensional variable selection, it is well known to exhibit a relatively high false positive rate and lead to biased estimates \citep{mazumder2011sparsenet} due to its over-penalization for larger coefficients. The group Minimax Concave Penalty (MCP) \citep{zhang2010nearly} is a non-convex penalty that reduces the estimation bias by gradually relaxing the penalization on coefficients as their magnitude increases \citep{breheny2015group}. MCP thus achieves a balance between sparsity and unbiased estimation, making it particularly attractive for variable selection in functional regression models \citep{chen2016variable,ghosal2020variable,ghosal2023variable}. Additionally, MCP is known to possess several desirable theoretical properties, such as the oracle property under standard regularity conditions \citep{zhang2010nearly,breheny2015group}, 
%meaning it can correctly identify the true model asymptotically and estimate the non-zero coefficients as efficiently as if the true model were known in advance. Furthermore, in both scalar and grouped settings, MCP has been shown to provide consistent variable selection and estimation, which 
This motivates its use in our VSFCOX method involving the coefficients corresponding to both scalar and functional covariates. The penalized estimation problem for the proposed VSFCOX method is formulated as:

% LASSO generally exhibits a higher false positive rate \citep{mazumder2011sparsenet}. Specifically, we propose using the group Minimax Concave Penalty (MCP) \citep{zhang2010nearly}, which mitigates the high bias issue of LASSO by gradually reducing the penalization rate as the coefficient magnitude increases \citep{breheny2015group}. MCP has been proven to ensure both selection and estimation consistency under standard assumptions in scalar regression. It also satisfies the oracle property, meaning that asymptotically, it performs like the oracle maximum likelihood estimator. These properties motivate our adoption of group MCP for functional variable selection.

% We apply group MCP on the coefficients and estimate \(\boldsymbol{\beta}\) and \(\boldsymbol{\gamma}\) as follows:
\begin{equation}
    (\hat{\boldsymbol{\beta}}, \hat{\boldsymbol{\gamma}})=\underset{\boldsymbol{\beta},\boldsymbol{\gamma}}{argmin} \left\{\frac{1}{n}(-2\ell(\boldsymbol{\beta},\boldsymbol{\beta}(\cdot))+\sum_{j=1}^{p} P_{MCP,\lambda,\phi} (\|\beta_{j}\|_{1})+\sum_{k=1}^K P_{MCP,\lambda,\phi} (\|\boldsymbol{\gamma_k}\|_{2}))\right\},
\end{equation}
where $P_{MCP,\lambda,\phi}(|\cdot|)$ is defined as :
\begin{equation*}
P_{MCP,\lambda,\phi}(|\cdot|)=
\begin{cases}
\lambda|\cdot|-\frac{|\cdot|}{2\phi}\hspace{1.4 cm} &\text{if $|\cdot|\leq \lambda\phi$}.\\
.5\lambda^2\phi \hspace{2.7 cm} &\text{if $|\cdot|>\lambda\phi$}.
\end{cases}
\end{equation*}
The penalized estimation criterion (6) for the VSFCOX method can be optimized using a group descent algorithm \citep{breheny2015group}. The details of this algorithm are presented in Web Appendix A. We have used the \texttt{grpreg} package \citep{breheny2015group} in R for performing the above optimization.

\subsection{Adaptive Penalized Estimation}
We also explore an adaptive version of our proposed penalty function based on adaptive weights for the penalized estimation criterion. Similar to the adaptive LASSO \citep{zou2006adaptive}, we define an adaptive penalization scheme \citep{gertheiss2013variable} by introducing weights \(w_k\) and \(v_k\) in the penalty function \ref{eq:6}. Specifically, the adaptive penalization approach is based on the penalty:
$
P_{\lambda,\psi}\{\beta_k (\cdot)\} = \lambda \left\{ w_k \int \beta_k (t)^2 \, dt + \psi v_k \int \beta_k''(t)^2 \, dt \right\}^{1/2},
$
where the weights \(w_k\) and \(v_k\) are chosen in a data-adaptive manner \citep{meier2009high}. These weights are designed to reflect subjective beliefs about the true parameter functions and allow for different levels of shrinkage and smoothness for various covariates. We then use a group-MCP generalization of the adaptive penalty as in the penalized estimation criterion (6). Denoting the initial estimated coefficient functions \(\beta_k (\cdot)\) by \(\acute{\beta_k (\cdot)}\), (using any fast methods)
%using the approach implemented in the R package {\tt mgcv}. 
the adaptive weights can be defined as \(w_k = 1/\|\acute{\beta_k (\cdot)}\|\) and \(v_k = 1/\|\acute{\beta_k'' (\cdot)}\|\), where $\|\acute{\beta_k (\cdot)}\|=\sqrt{\int \acute{\beta_k (t)^2} \, dt}$ is the $\mathscr{L}^2$ norm of $\acute{\beta_k (\cdot)}$ (and similar for $\acute{\beta_k'' (\cdot)}$). 

%Adaptive estimation has been shown to significantly reduce the number of false positives in penalty-based variable selection \citep{meier2009high,gertheiss2012regularization}. %Given that the computation of the initial estimates is not time-consuming, the computational burden for the adaptive penalty is comparable to that of the standard (non-adaptive) penalty.

\subsection{Choosing the tuning parameters}
Until now, we have assumed that the sparsity parameter \(\lambda\) and the smoothness parameter \(\psi\) in the penalties were known. To determine the optimal tuning parameters \(\psi\) (for smoothness) and \(\lambda\) (for sparsity), we employ an extended Bayesian information type criterion (EBIC) \citep{chen2008extended}. The proposed EBIC is defined as $EBIC(\lambda,\psi) = BIC + 2 \log{p \choose \nu}$, where we denote by $\nu$ the number of selected scalar and functional variables, and $p$ denotes the total number of scalar and functional variables. In the context of functional data,  we have used this approach for calculating $\nu$ based on the number of selected functional variables rather than the total number of nonzero basis coefficients. The optimal value of $(\lambda,\psi)$ is then chosen using a two-dimensional grid search, producing the minimum EBIC.
%This approach introduces a key departure from traditional approaches by employing the concept of group variables rather than individual variables when calculating the number of selected variables.
% To clarify this approach with an example, consider a fitted model where there are 10 group variables, each containing 5 individual variables. Suppose only 2 groups are selected, meaning all individual variables in the non-selected groups are set to 0. Within the selected groups, 2 individual variables are non-zero in one group, and 3 individual variables are non-zero in the other. Traditional methods for calculating degrees of freedom would sum the non-zero individual variables, yielding a total of $2+3=5$. In contrast, our method focuses on the effective number of selected groups, which in this case is 2, regardless of the number of non-zero individual variables within each group. This group-based perspective provides a more accurate and parsimonious estimation, as demonstrated through our simulation study.
For the tuning parameter \(\phi\), we use the value \(3\) for the MCP, following the recommendation of the original authors \citep{zhang2010nearly}. 
%\vspace{- 7 mm}

\section{Simulation Study}
\label{sec:sim_stud}
\subsection{Simulation Setup}
In this section, we assess the effectiveness of our variable selection method, VSFCOX, through a simulation study. We simulate data from the following FLCM,
\begin{equation*}
\log h_i(t; \boldsymbol{\beta}, \boldsymbol{\beta}(\cdot)) = \log h_0(t) + \boldsymbol{Z_{i}}^T \boldsymbol{\beta} + \sum_{k=1}^{20} \int_0^1 Z_{ik}(s) \beta_k(s) \, ds,
\end{equation*}
where $\boldsymbol{\beta}=(\beta_1, \beta_2, \cdots, \beta_{15})^T \in \mathbb{R}^{15}$, and $\boldsymbol{Z_{i}}=( Z_{i1}, Z_{i2}, \ldots, Z_{i15} )^T \in \mathbb{R}^{15}$. The regression parameters are specified as \(\beta_{1} = 1\), \(\beta_{2} = 1.5\), and \(\beta_{3} = 2\), with the remaining parameters \(\beta_j = 0\) for \(j = 4, 5, 6, \ldots, 15\), indicating that the last 12 scalar covariates are not relevant. We denote by $S=\{m/100: m=0,1,\ldots, M=100\}$ the grid of time points in $[0,1]$, over which the functional curves are observed. The regression functions are given by \(\beta_{1}(s) = 3\cos(\pi s)\), \(\beta_{2}(s) = 4.5\sin(\pi s)\), \(\beta_{3}(s) = 3.5\cos(2\pi s) - 5.5\sin(2\pi s)\), \(\beta_{4}(s) = 4\cos(2\pi s)\), and \(\beta_{5}(s) = 2.5\sin(2\pi s)\), with the remaining functions \(\beta_k(s) = 0\) for \(k = 6, 7, 8, \ldots, 20\), indicating that the last 15 functional covariates are not relevant. The scalar covariates \(Z_{ij} \stackrel{\text{iid}}{\sim} Z_j\), where \(Z_j \sim \text{Uniform}(-1, 1)\). The functional covariates \(\{Z_k(s), k = 1, 2, \ldots, 20\}\) are generated as \(Z_{ik}(s) = \sum_{q = 1}^{20} \omega_{ik_q} \phi_q(s)\), where \(\phi_q(s)\) are orthogonal basis polynomials and \(\omega_{ik_q}\) are mean zero, independently normally distributed scores with variance \(\sigma^2_q = 4q\). The baseline hazard follows an exponential distribution with a rate \(\exp(0.5)\), resulting in $log\{h_0 (t)\}=\beta_0 = 0.5$. The censoring times are drawn independently from an exponential distribution with a rate parameter of $\frac{1}{\mu_c}$, where the mean censoring time, $\mu_c=10$. We consider three different sample sizes: \(n \in \{200, 400, 800\}\). For each sample size, we utilize $n_d =200$ Monte Carlo (MC) replications to evaluate our method.

\subsection{Simulation results}
Our primary focus is on selecting the relevant scalar covariates \(Z_1, Z_2, Z_3\) and functional covariates \(Z_1(s), Z_2(s), Z_3(s), Z_4(s), Z_5(s)\). Additionally, we aim to accurately estimate both the scalar effects \(\beta_1, \beta_2, \beta_3\) and the functional parameter curves \(\beta_1(s), \beta_2(s), \beta_3(s), \beta_4(s), \beta_5(s)\). We use a cubic B-spline basis with 10 basis functions to model the regression functions $\beta_k (s)$. We compare the performance of the proposed VSFCOX against an approach using group LASSO penalty (grpregLASSO). The tuning parameters are automatically selected using the proposed EBIC, incorporating both sparsity and smoothness. Table \ref{t1} presents the variable selection performance. Specifically, it reports the true positive rate (TPR) and false positive rate (FPR) for scalar, functional, and all variables combined across three sample sizes. Additionally, the table displays the average model size for each scenario. From the results, it is evident that VSFCOX consistently performs well in terms of selection accuracy, accurately selecting the relevant variables and discarding the irrelevant ones. Both VSFCOX and grpregLASSO demonstrate a high TPR, indicating perfect recovery of true signals. However, only VSFCOX maintains a low FPR, with zero false positives for functional covariates and a negligible FPR for the scalar covariates, highlighting its robustness in eliminating the irrelevant features. 
%The FPR of our proposed method consistently decreases as the sample size increases. Furthermore, VSFCOX produces stable average model sizes across different sample sizes, effectively balancing parsimony and accuracy. 
Overall, the results underscore the efficacy of VSFCOX in both scalar and functional variable selection, particularly as sample size increases.

\begin{table}[H]
\centering
\caption{Comparison of selection performance and average model size across different sample sizes}
\label{t1}
\begin{tabular}{cccccc}
\hline
Sample size & Method      & Variables  & TPR   & FPR   & Average model size      \\ \hline
200         & grpregLASSO & scalar     & 0.993 & 0.920 & \multirow{3}{*}{32.965} \\
            &             & functional & 0.993 & 0.932 &                         \\
            &             & all        & 0.993 & 0.927 &                         \\
            &             &            &       &       &                         \\
            & VSFCOX      & scalar     & 0.990 & 0.033 & \multirow{3}{*}{8.355}  \\
            &             & functional & 0.999 & 0     &                         \\
            &             & all        & 0.996 & 0.014 &                         \\
            &             &            &       &       &                         \\
400         & grpregLASSO & scalar     & 1     & 0.265 & \multirow{3}{*}{13.95}  \\
            &             & functional & 1     & 0.185 &                         \\
            &             & all        & 1     & 0.220 &                         \\
            &             &            &       &       &                         \\
            & VSFCOX      & scalar     & 1     & 0.003 & \multirow{3}{*}{8.030}  \\
            &             & functional & 1     & 0     &                         \\
            &             & all        & 1     & 0.001 &                         \\
            &             &            &       &       &                         \\
800         & grpregLASSO & scalar     & 1     & 0.308 & \multirow{3}{*}{15.235} \\
            &             & functional & 1     & 0.236 &                         \\
            &             & all        & 1     & 0.268 &                         \\
            &             &            &       &       &                         \\
            & VSFCOX      & scalar     & 1     & 0     & \multirow{3}{*}{8.005}  \\
            &             & functional & 1     & 0     &                         \\
            &             & all        & 1     & 0     &                         \\ \hline
\end{tabular}
\end{table}

Beyond selection performance, VSFCOX also demonstrates strong performance in estimation accuracy. It consistently yields minimal bias and mean squared error (MSE) for the scalar parameters (see Web Table 1) and low mean integrated squared error (MISE) for the functional parameters reported in Table \ref{t3p1}. The MISE of the functional estimate \(\beta_k(s)\) is defined as:  
\[
MISE_k = \frac{1}{n_d}\sum_{d=1}^{n_d} \left( \int_0^1 \left( \hat{\beta}_{kd}(s) - \beta_k(s) \right)^2 \, ds \right),
\]
where \(\hat{\beta}_{kd}(s)\) represents the estimate of \(\beta_k(s)\) for the \(d\)-th generated dataset, and \(n_d\) is the total number of datasets. 
%This measure captures the average discrepancy between the estimated and true functional coefficients over multiple replicates, integrated across the entire domain. 
As the sample size increases, VSFCOX's bias and MSE (and MISE for functional coefficients) steadily decrease, demonstrating its efficiency in the estimation of the scalar and functional coefficients.

\begin{table}[h!]
\centering
\caption{Comparison of MISE across different sample sizes}
\label{t3p1}
\begin{tabular}{ccccccc}
\hline
Sample size & $\hat{\beta_1}(\cdot)$ & $\hat{\beta_2}(\cdot)$ & $\hat{\beta_3}(\cdot)$ & $\hat{\beta_4}(\cdot)$ & $\hat{\beta_5}(\cdot)$ \\ \hline
200             & 1.168               & 1.994               & 3.569               & 1.614               & 0.984               \\
            &        &                     &                     &                     &                     &                     \\
400             & 0.221               & 0.321               & 0.543               & 0.289               & 0.179               \\
            &        &                     &                     &                     &                     &                     \\
800             & 0.067               & 0.085               & 0.136               & 0.084               & 0.055               \\ \hline
\end{tabular}
\end{table}

We display the Monte Carlo mean of the estimated functional coefficients \(\hat{\beta}_k(\cdot)\) (\(k=1,2,3,4,5\)), using our proposed method overlaid on true regression curves in Figure \ref{figs} for \(n=400\). We also display the Monte-Carlo point-wise 95\% confidence intervals (CIs). It can be observed that the VSFCOX method closely tracks the true curves, effectively capturing the underlying functional associations. The results for sample sizes $n=200$ and $n=800$, which demonstrate similar patterns are displayed in Web Figures 1 and 2. With the increased sample size, the estimates show improved accuracy and reduced variability, as reflected by narrower confidence intervals. Overall, these results highlight that VSFCOX not only excels in variable selection but also provides highly accurate estimates of both scalar and functional parameters. 
%Its ability to maintain low bias, low MSE, and minimal MISE as the sample size increases underscores its robustness and effectiveness.

\newcommand{\solidline}{\raisebox{2pt}{\tikz{\draw[-,black,solid,line width = 0.3pt](0,0) -- (5mm,0);}}}

\begin{figure}[h!]
    \centering
    \includegraphics[scale=0.8]{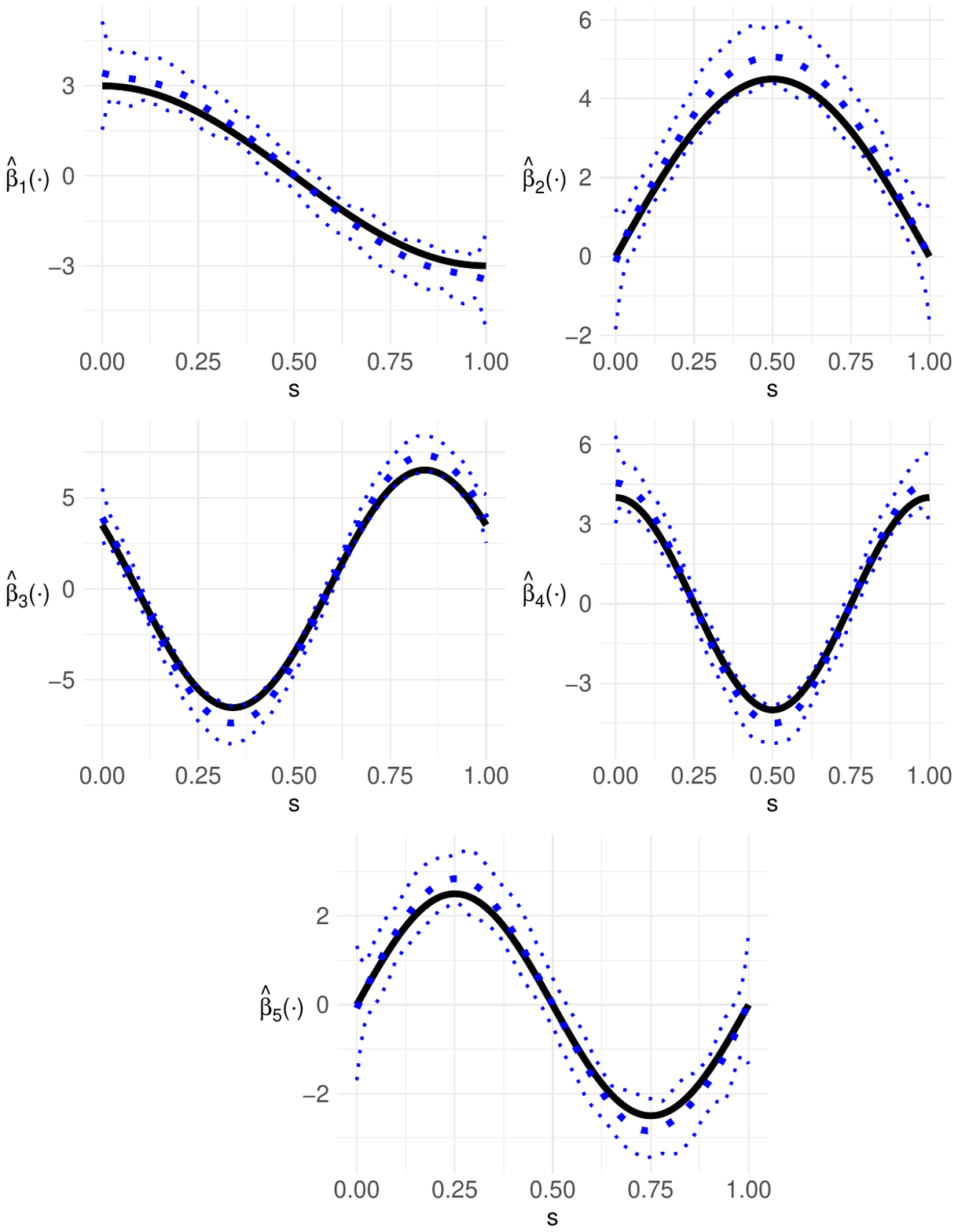}
    \caption{MC estimates and pointwise confidence intervals of the coefficient functions ($n=400$); (\textcolor{blue}{\textbf{$\cdots$}}, VSFCOX; \protect\solidline, true curve)}
    \label{figs}
\end{figure}

% \begin{figure}[h!]
%     \centering
%     \begin{subfigure}{0.49\textwidth}
%         \centering
%         \includegraphics[width=\textwidth]{bt1.pdf}
%         \caption{}
%         \label{fig:plot1}
%     \end{subfigure}
%     \hfill
%     \begin{subfigure}{0.49\textwidth}
%         \centering
%         \includegraphics[width=\textwidth]{bt2.pdf}
%         \caption{}
%         \label{fig:plot2}
%     \end{subfigure}
%     \vspace{0.5cm} % Adjust spacing between rows
%     \begin{subfigure}{0.49\textwidth}
%         \centering
%         \includegraphics[width=\textwidth]{bt3.pdf}
%         \caption{}
%         \label{fig:plot3}
%     \end{subfigure}
%     \hfill
%     \begin{subfigure}{0.49\textwidth}
%         \centering
%         \includegraphics[width=\textwidth]{bt4.pdf}
%         \caption{}
%         \label{fig:plot4}
%     \end{subfigure}
%     \vspace{0.5cm} % Adjust spacing between rows
%     \begin{subfigure}{0.49\textwidth}
%         \centering
%         \includegraphics[width=\textwidth]{bt5.pdf}
%         \caption{}
%         \label{fig:plot5}
%     \end{subfigure}

%     \caption{MC estimates and pointwise confidence intervals of the coefficient functions ($n=400$); (\textcolor{blue}{\textbf{$\cdots$}}, VSFCOX; \protect\solidline, true curve);(a) $\beta_1(\cdot)$; (b) $\beta_2(\cdot)$; (c) $\beta_3(\cdot)$; (d) $\beta_4(\cdot)$; (e) $\beta_5(\cdot)$}
%     \label{figs}
% \end{figure}

We also explored adaptive penalized estimation, which demonstrates marginal improvements in the selection performance for smaller sample sizes. We refer to Web Tables 2-4 and Web Figure 3 in the supporting information for the detailed results.

\section{Real data application: Modelling All-Cause Mortality in NHANES 2003-2006}
\label{realdat}
We apply the VSFCOX method to accelerometer data from the NHANES waves 2003-2006, to identify the key temporally varying distributional patterns of physical activity and demographic, lifestyle predictors associated with all-cause mortality. While numerous studies have established a consistent link between higher levels of PA and reduced mortality risk, majority have relied on scalar summary metrics for PA, such as total activity count (TAC) or moderate-to-vigorous PA (MVPA) \citep{varma2017re,yerramalla2021association,ledbetter2022cardiovascular}. Although these summary-metric-based approaches simplify interpretation, they cannot capture the full spectrum of changes in PA intensities over time \citep{ghosal2023distributional}. Since PA trends can fluctuate throughout the day, preserving these temporal details is crucial for understanding the circadian rhythm of PA \citep{xiao2015quantifying} and its implications for all-cause mortality \citep{cui2021additive}. Recent research \citep{ghosal2022scalar} has also demonstrated that distributional metrics beyond mean PA, such as higher-order moments like variability and skewness, offer valuable and complementary insights to those provided by mean PA alone. The daily time-of-day dependent L-moments \citep{ghosal2022scalar, ghosal2023distributional,cho2024exploring} provide a useful framework for characterizing the daily distributional patterns of physical activity beyond the average diurnal PA pattern. We are interested in identifying the key daily distributional patterns of PA related to all-cause mortality among older US adults.

%It is essential to effectively leverage the time-varying distributional patterns within the data. Considering both temporal and distributional patterns simultaneously may help capture unique behavioral patterns \citep{ghosal2022scalar}. Temporally varying distributional information can be modeled as functional observations over a 24-hour period using FDA approaches \citep{ghosal2022scalar,cho2024exploring}. Given the complexity of such data, variable selection is essential to identify key PA patterns while maintaining model interpretability and parsimony. 
%Applying our variable selection method to the NHANES dataset, we aim to identify critical scalar and functional covariates that influence all-cause mortality among older US adults.

The National Health and Nutrition Examination Survey (NHANES) provides comprehensive health and nutrition statistics for the civilian non-institutionalized US population. NHANES 2003-06 includes objectively measured physical activity data collected by hip-worn accelerometers. Participants were asked to wear a physical activity monitor starting on the day of their exam and to keep wearing it all day and night for seven full days (midnight to midnight), removing it on the morning of the ninth day \citep{leroux2019organizing}. NHANES data is linked to the National Death Index (NDI) for collecting mortality information. Specifically, we use the 2019 mortality data (as of December 31) from the NDI (\url{https://www.cdc.gov/nchs/data-linkage/mortality-public.htm}) to define our survival outcome.

Our final sample consists of a total of 2,816 adults aged 50–85 years, who had physical activity monitoring data available for at least ten hours per day over a minimum of four days \citep{cui2021additive,ghosal2023functional}. Survival time is measured in years from the end of accelerometer wear, with all subjects censored on December 31, 2019, based on mortality information from the NDI 2019 release. Of the 2816 study participants at baseline, 1117 (39.7\%) were deceased, while the remaining 1699 participants were considered right-censored. Web Table 5 presents the descriptive statistics of the complete sample and also stratified by deceased or survivor.
The average follow-up time for participants was 12.4 years. Web Figure 4 displays the distribution of the observed survival times and the Kaplan-Meier estimate. The scalar covariates included in our analysis were age ($Z_{1}$), BMI ($Z_{2}$), gender ($Z_{3}$), and smoking status (never (reference), former ($Z_{4}$) and current ($Z_{5}$)). 
%Web Table 1 presents the baseline demographic characteristics of the full study cohort, as well as those of participants categorized by survival status.

% \begin{figure}[h!]
%     \centering
%     \includegraphics[scale=0.45]{combHK.pdf}
%     \caption{Histogram of survival time and Kaplan-Meier marginal survival curve}
%     \label{fig4}
% \end{figure}

% \begin{figure}[h!]
%   \centering
%     \begin{subfigure}{0.45\textwidth}
%         \centering
%         \includegraphics[scale=0.45]{Hist.pdf}
%         \caption{Histogram of survival time}
%     \end{subfigure}
%     \hfill
%         \begin{subfigure}{0.45\textwidth}
%         \centering
%         \includegraphics[scale=0.45]{KM.pdf}
%         \caption{Kaplan-Meier marginal survival curve}
%     \end{subfigure}
% \caption{Histogram of survival time and Kaplan-Meier marginal survival curve}
% \label{fig4}
% \end{figure}

We calculate daily time-of-day dependent L-moments to serve as our functional exposures, effectively capturing the time-varying distributional patterns in the PA data \citep{ghosal2022scalar}. L-moments are robust rank-based analogues of traditional moments \citep{hosking1990moments,ghosal2023distributional}.The sample L-moments are linear combinations of order statistics (L-statistics), and can be used to compute quantities analogous to the mean, standard deviation, skewness, and kurtosis, termed $L_1$ moment (which coincides with mean), L-scale ($L_2$), L-skewness ($L_3$), and L-kurtosis ($L_4$), respectively. In particular, the \( r \)-th order population L-moment of a random variable \( X \) is defined as:
\begin{equation*}
    L_r = r^{-1} \sum_{k=0}^{r-1} (-1)^k {\binom{r-1}{k}} E(X_{r-k:r}) \hspace{3mm} r=1,2,\ldots,
\end{equation*}
where \( X_{1:n} \leq X_{2:n} \leq \ldots \leq X_{n:n} \) denote the order statistics of a random sample of size \( n \) drawn from the distribution of \( X \).

For each participant \(i = 1, \dots, n\), we denote the log-transformed minute-level physical activity counts for individual \(i\) on day \(j\) at time \(s\) as  \(X_{ij}(s)\) for \(j = 1, \dots, n_i\).
The observation times are given by are $S=\{m/60: m\in {0,1,2,\ldots,1439}\}$ hours. To capture the daily distributional patterns of physical activity data, we extract the first four diurnal L-moments profiles \(L_{ik}(s)\) (\(k=1,2,3,4\)) \citep{ghosal2022scalar}, where \(L_{ir}(s)\) represents the \(r\)-th L-moment of \(\{X_{ij}(s)\}_{j=1}^{n_i}\) for \(s \in (s-\zeta,s+\zeta)\), with $\zeta=\frac{5}{60}$ hour (or 5 minutes). The average smoothed L-moments are displayed in Figure \ref{1b} along with the profiles for five randomly selected participants. We define \(Z_{ik}(s) = L_{ik}(s)\) for \(k = 1,2,3,4\). Additionally, to explore the effect modification by age and gender in the effect of the diurnal L-moments, we incorporate the following interaction terms: \(Z_{ik}(s)\) for \(k = 5,6,7,8\) which are the interactions between age and the first four L-moments; and \(Z_{ik}(s)\) for \(k = 9,10,11,12\) representing the interactions between gender and the first four L-moments. We restrict the focus of our analysis to hours \(s \in [6, 22]\), corresponding to the time range from 6 a.m. to 10 p.m. since most people are inactive during the night \citep{ghosal2023variable} and report zero activity in NHNAES 2003-06 outside this period. Finally, we have 5 scalar covariates and 12 functional covariates in our functional linear Cox model given by:
\begin{equation*}
\log h_i(t| Z_{i1},\ldots, Z_{i5}, Z_{i1}(\cdot), \ldots, Z_{i12}(\cdot)) = \log h_0(t) + \sum_{j=1}^{5} Z_{ij}\beta_j + \sum_{k=1}^{12} \int_{6}^{22} Z_{ik}(s) \beta_k(s) \, ds.
\end{equation*}

% \begin{figure}[h!]
%     \centering
%     \includegraphics[scale=0.45]{combLm.pdf}
%     \caption{The average of the first four L-moments of the time-varying physical activity profiles across all participants, as well as the first four L-moments for five randomly selected participants by SEQN, using log-transformation}
%     \label{fig3}
% \end{figure}

\begin{figure}[h!]
    \centering
    \includegraphics[scale=0.45]{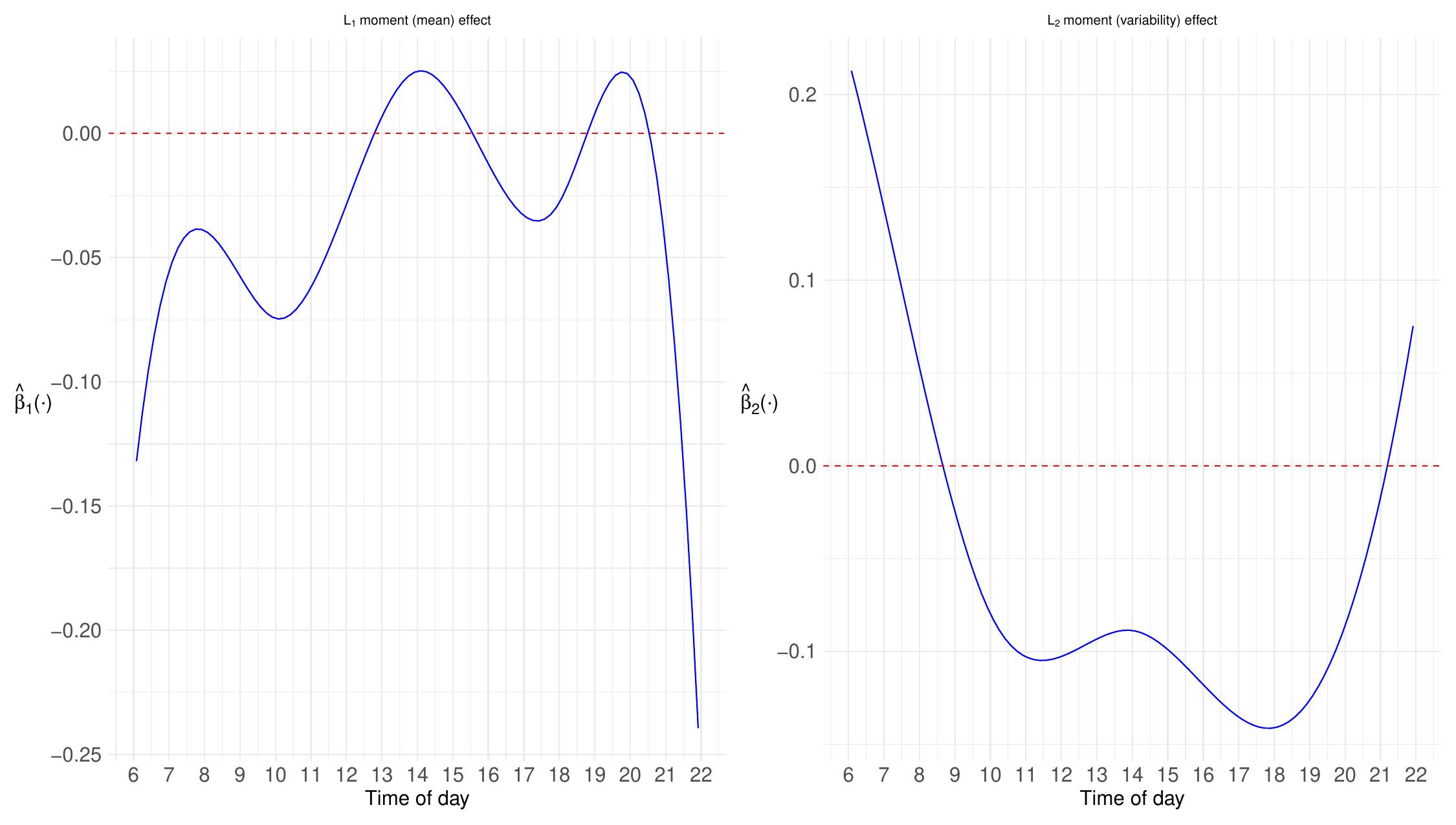}
    \caption{Estimated functional effects of the first ($L_1$) and second ($L_2$) moments over time from 6 a.m. to 10 p.m.}
    \label{fig5}
\end{figure}

We applied the proposed VSFCOX method to the above FLCM for selection and estimation of the functional and scalar coefficients. For the functional variables, only the first two L-moments were selected, while all others, including interaction terms, were excluded. Figure \ref{fig5} displays the estimated functional coefficients $\hat{\beta}_1(\cdot),\hat{\beta}_2(\cdot)$ of the the first and second diurnal L-moments (\(L_1(\cdot)\) and \(L_2 (\cdot)\)) of PA. The first diurnal L-moment represents the average diurnal PA, while the second diurnal L-moment captures the variability of PA throughout the day. For \(\hat{\beta}_1(s)\), associated with the first diurnal L-moment, a higher daily mean PA during both the morning period (6 a.m. to 1 p.m.) and dusk (approximately 4 p.m. to 7 p.m.) is found to be associated with a reduced hazard of all-cause mortality. This finding highlights the protective effect of a higher average physical activity during these key periods of the day \citep{leroux2019organizing,cui2021additive,leroux2024nhanes}. For \(\hat{\beta}_2(s)\), associated with the second diurnal L-moment, a higher daily variability in PA \((L_2(\cdot))\) between approximately 9 a.m. and 9 p.m. is found to be associated with a reduced hazard of all-cause mortality. These results suggest that not only the average diurnal PA, but also the variability of PA throughout the day could be an important protective factor against the risk of all-cause mortality \citep{cho2024exploring} among older adults, possibly leading to a higher reserve of PA. 

Web Table 6 displays the selected scalar predictors along with their estimated effects in the form of hazard ratios. The selected scalar variables are age, gender (female), current smoking, and former smoking. Specifically, age \citep{leroux2024nhanes}, current smoking, and former smoking \citep{jha201321st} are found to be associated with an increased hazard of all-cause mortality. The hazard associated with current smoking exceeded that of former smoking, underscoring the severe harm of ongoing smoking \citep{kenfield2008smoking}. Females are found to have a lower hazard of all-cause mortality \citep{haukkala2009gender} in this study than males, after adjusting for all other predictors.

In additional analysis, we introduced 10 functional pseudo-covariates \citep{wu2007controlling,ghosal2020variable}  to the original data to rigorously assess the performance of the proposed variable selection approach. These functional pseudo-covariates serve as ``noise" to test our method’s specificity and its potential for false-positive selection in high-dimensional case. Specifically, we generate $Z_{ij}(\cdot)\stackrel{iid}{\sim} Z_j(\cdot)$, where $Z_j(h)$ ($j=13,14,\ldots,22$) are given by $Z_j(h)=a_j\hspace{1mm}\sqrt[]{2}sin(\pi j h/24)+ b_j\hspace{1mm}\sqrt[]{2}cos(\pi j h/24)$ with $a_j\sim \mathcal{N}(0,(2)^2)$, $b_j\sim \mathcal{N}(0,(2)^2)$. In total, we have 27 covariates, where the first 17 are the original covariates, and the remaining 10 are simulated predictors. We then apply the VSFCOX method to observed survival time and censoring indicator \( \{Y_i, \delta_i\} \), and the covariates \( Z_{i1}, Z_{i2}, Z_{i3}, \dots, Z_{i1}(s), Z_{i2}(s), \dots, Z_{i22}(s) \). This process is repeated a large number ($B=100$) times to observe which variables are selected in each iteration. The selection percentages of the variables are reported in Table \ref{t4}. 
%We expect our method to consistently identify the truly influential predictors while disregarding the randomly generated functional covariates in most cases.

\begin{table}[h!]
\centering
\caption{Selection percentages of the variables in the NHANES application.}
\label{t4}
\resizebox{\textwidth}{!}{\begin{tabular}{cccccccc}
\hline
Age & Gender (Female) & Former Smoker & Current Smoker & $L_1$  & $L_2$  & Remaining Original Variables & Pseudo Variables \\ \hline
100\% & 100\%             & 100\%           & 100\%            & 100\% & 100\% & 0\%                            & 0\%                \\ \hline
\end{tabular}}
\end{table}
Variables such as age, gender (female), former Smoker, current Smoker, and functional predictors diurnal $L_1 (\cdot)$ moment and the diurnal $L_2 (\cdot)$ moment were selected 100\% of the time, demonstrating their strong influence on all-cause mortality. In contrast, the remaining original variables and pseudo variables had a selection percentage of 0\%, suggesting that they were not deemed relevant in the selection process. This result further highlights the effectiveness of the VSFCOX method in identifying the truly influential predictors while eliminating the irrelevant ones. %Notably, our findings highlight that both the daily mean ($L_1$ moment) and daily variability ($L_2$ moment) of physical activity are significantly associated with the hazard of all-cause mortality in US older people, aligning with the results of \cite{cho2024exploring}.

\section{Discussion}
\label{disc}
In this paper, we proposed a variable selection method in the functional linear Cox model with multiple functional and scalar covariates. Our proposed methodology, VSFCOX, integrates penalized
spline-based estimation with a group minimax concave type penalty (MCP), enforcing smoothness and sparsity into the estimation of the functional coefficients. An efficient group descent algorithm is employed for the optimization, and an automated selection criterion is provided for the choice of the tuning parameters.

%estimation techniques, particularly group MCP, to efficiently identify influential predictors in high-dimensional survival models. Through extensive simulation studies and real-world applications using NHANES data, we demonstrated the effectiveness of our approach in selecting relevant covariates while integrating smoothness into the estimation of functional coefficients.

The simulation results highlighted a satisfactory selection and estimation accuracy of VSFCOX, showing nearly perfect selection consistency for both scalar and functional covariates with increasing sample sizes. 
%Additionally, VSFCOX demonstrated lower bias, MSE, and MISE in parameter estimation, confirming its robustness in functional variable selection. The Monte Carlo estimates of the coefficient functions further reinforced these findings, with VSFCOX estimates closely aligning with the true underlying functional curves.
The application of our method to the NHANES 2003-06 dataset provided valuable epidemiological insights into the association between daily distributional patterns of PA and all-cause mortality among older US adults. VSFCOX identified the first and second diurnal L-moments (\(L_1(\cdot)\) and \(L_2 (\cdot)\)) of PA, in addition to age, gender, and smoking status, as the key drivers of all-cause mortality among older adults. The results indicate that in addition to daily average PA, daily variability in PA is also a significant determinant of all-cause mortality, after adjusting for age, gender, BMI, and smoking status. These results can be helpful for designing time-of-day \citep{feng2023associations} and intensity-specific PA interventions \citep{cho2024exploring}.
%This demonstrates the practical utility of our selection technique in handling complex survival data with multiple functional and scalar predictors and can help design time-of-day and intensity-specific PA interventions based on different demographic factors.

In this article, we have developed VSFCOX specifically for the functional linear Cox Model. An important future direction would be relaxing the assumption of linearity in covariate effects. In many real-world applications, predictor effects may exhibit nonlinear or more complex patterns that linear models cannot adequately capture. The proposed variable selection method could be extended to accommodate additive models \citep{cui2021additive}
which would offer greater flexibility by allowing nonlinear additive effects of the functional and scalar predictors. Sparse single-index models \citep{jiang2011functional,ma2016estimation,ghosal2024variable} would be another interesting direction to explore, going beyond additive models. If the proportional hazards assumption is not suitable, the variable selection method can be extended to more flexible survival models such as the functional time-transformation model \citep{ghosal2025functional}.

In this paper, the primary focus has been on variable selection and estimation of the functional effects. A key challenge in high-dimensional models is that standard inferential tools are no longer directly valid following variable selection using regularized techniques. Future research will therefore explore developing uncertainty quantification methods for the estimated functional effects in VSFCOX using post-selection inference approaches \citep{lee2016exact,taylor2018post} and sample-splitting strategies \citep{wasserman2009high} that can mitigate the dependency between selection and inference stages.

\section*{Acknowledgement}
This work is supported by the SPARC Graduate Research grant from the Office of the Vice President for Research, University of South Carolina.

\vspace*{- 7 mm}
\section*{Data Availability Statement}
The data supporting the findings of this study are 
publicly available at \url{https://wwwn.cdc.gov/nchs/nhanes/continuousnhanes/}.

%accessible through the \texttt{rnhanesdata} R package \citep{rnh}, available at \url{https://github.com/andrew-leroux/rnhanesdata}.

\bibliographystyle{biom}
\bibliography{score.bib}

\vspace{- 6 mm}
\section*{Supporting Information}
Web Tables 1-6 and Web Figures 1-4 referenced in this article are available with this paper at the Biometrics website on Wiley Online Library. Software illustration of the proposed method is provided with this paper and also will be available online at Github. %\url{https://github.com/rahulfrodo/GFOSR_Selection}.

%\section*{Software}
%Software illustrating implementation of the proposed method is available with this article at the Biometrics website on Wiley Online Library.
\end{document}

\maketitle
%Web Algorithm 1, Web Figure 1-4, and Web Tables 1-6 referenced in the paper are given below.

\section{Web Appendix A}
\begin{algorithm}[H]
\caption{Group descent algorithm for the VSFCOX method}
\begin{algorithmic}
\State Initialize $\boldsymbol{\beta}^{(0)}, \boldsymbol{\gamma}^{(0)}$
\Repeat
    \State Compute linear predictor: $\boldsymbol{\eta} \leftarrow \boldsymbol{Z} \boldsymbol{\beta} + \sum_{k=1}^{K} \tilde{\boldsymbol{Z}}_k^T \boldsymbol{\gamma}_k$
    \State Compute working residual: $\tilde{\boldsymbol{r}} \leftarrow \boldsymbol{\Delta}-\boldsymbol{e}$
    \State \textbf{for} $j = 1, 2, \dots, p$
    \State \hspace{4mm} $z_j \leftarrow \boldsymbol{Z}_j^T \tilde{\boldsymbol{r}} + \beta_j$
    \State \hspace{4mm} $\beta_j' \leftarrow F\left(|z_j|, \lambda, \phi\right) \frac{z_j}{|z_j|}$ \Comment{Firm-thresholding}
    \State \hspace{4mm} $\tilde{\boldsymbol{r}}' \leftarrow \tilde{\boldsymbol{r}} - \boldsymbol{Z}_j (\beta_j' - \beta_j)$
    \State \textbf{for} $k = 1, 2, \dots, K$
    \State \hspace{4mm} $\boldsymbol{z}_k \leftarrow \tilde{\boldsymbol{Z}}_k^T \tilde{\boldsymbol{r}} + \boldsymbol{\gamma}_k$
    \State \hspace{4mm} $\boldsymbol{\gamma}_k' \leftarrow F(\|\boldsymbol{z}_k\|, \lambda, \phi)$
    \State \hspace{4mm} $\tilde{\boldsymbol{r}}' \leftarrow \tilde{\boldsymbol{r}} - \tilde{\boldsymbol{Z}}_k (\boldsymbol{\gamma}_k' - \boldsymbol{\gamma}_k)$
\Until{convergence}
\end{algorithmic}
\end{algorithm}

\hspace*{ 2 mm}
Expected event: \[e_i=\sum_{j : i \in R_j} \delta_j \frac{e^{\eta_i}}{\sum_{k \in R_j} e^{\eta_k}}.\]

Firm-thresholding: \[F\left(z, \lambda, \phi\right)=\begin{cases}
\displaystyle \frac{S(z, \lambda)}{1 - 1/\phi} & \text{if } |z| \leq \lambda \phi \\
z & \text{if } |z| > \lambda \phi 
\end{cases}.\]

Soft-thresholding: \[S(z, \lambda)=\begin{cases}
\displaystyle z-\lambda & \text{if } z > \lambda \\
0 & \text{if } |z| \leq \lambda \\
z+\lambda & \text{if } z < -\lambda 
\end{cases}.\]

\newpage

\section{Web Tables}
\begin{table}[h!]
\centering
\caption{Comparison of bias and MSE across different sample sizes}
\label{t2}
\begin{tabular}{ccccccccc}
\hline
Sample size & \multicolumn{2}{c}{$\hat{\beta_1}$} &  & \multicolumn{2}{c}{$\hat{\beta_2}$} &  & \multicolumn{2}{c}{$\hat{\beta_3}$} \\ \cline{2-3} \cline{5-6} \cline{8-9} 
            & Bias             & MSE              &  & Bias             & MSE              &  & Bias             & MSE              \\ \hline
200         & 0.364            & 0.300            &  & 0.539            & 0.452            &  & 0.747            & 0.709            \\
            &                  &                  &  &                  &                  &  &                  &                  \\
400         & 0.141            & 0.044            &  & 0.214            & 0.072            &  & 0.251            & 0.097            \\
            &                  &                  &  &                  &                  &  &                  &                  \\
800         & 0.049            & 0.011            &  & 0.090            & 0.017            &  & 0.100            & 0.025            \\ \hline
\end{tabular}
\end{table}

\begin{table}[h!]
\centering
\caption{Comparison of selection performance and average model size across different sample sizes - Adaptive Penalty}
\label{t1ad}
\begin{tabular}{ccccc}
\hline
Sample size & Variables  & TPR   & FPR   & Average model size     \\ \hline
200         & scalar     & 0.992 & 0.031 & \multirow{3}{*}{8.355} \\
            & functional & 1.000 & 0.000 &                        \\
            & all        & 0.997 & 0.014 &                        \\
            &            &       &       &                        \\
400         & scalar     & 1     & 0.003 & \multirow{3}{*}{8.035} \\
            & functional & 1     & 0     &                        \\
            & all        & 1     & 0.001 &                        \\
            &            &       &       &                        \\
800         & scalar     & 1     & 0     & \multirow{3}{*}{8}     \\
            & functional & 1     & 0     &                        \\
            & all        & 1     & 0     &                        \\ \hline
\end{tabular}
\end{table}

\begin{table}[h!]
\centering
\caption{Comparison of bias and MSE across different sample sizes - Adaptive Penalty}
\label{t2ad}
\begin{tabular}{ccccccccc}
\hline
Sample size & \multicolumn{2}{c}{$\hat{\beta_1}$} &  & \multicolumn{2}{c}{$\hat{\beta_2}$} &  & \multicolumn{2}{c}{$\hat{\beta_3}$} \\ \cline{2-3} \cline{5-6} \cline{8-9} 
            & Bias             & MSE              &  & Bias             & MSE              &  & Bias             & MSE              \\ \hline
% 100         & 1.263         & 6.709         &  & 2.560         & 13.728         &  & 3.631          & 24.868         \\
%             &                  &                  &  &                  &                  &  &                  &                  \\
200         & 0.345            & 0.268            &  & 0.553            & 0.447            &  & 0.778            & 0.768            \\
            &                  &                  &  &                  &                  &  &                  &                  \\
400         & 0.138            & 0.041            &  & 0.208            & 0.069            &  & 0.251            & 0.095            \\
            &                  &                  &  &                  &                  &  &                  &                  \\
800         & 0.051            & 0.011            &  & 0.087            & 0.017            &  & 0.105            & 0.025            \\ \hline
\end{tabular}
\end{table}

\begin{table}[h!]
\centering
\caption{Comparison of MISE across different sample sizes - Adaptive Penalty}
\label{t3ad}
\begin{tabular}{ccccccc}
\hline
Sample size & $\hat{\beta_1}(\cdot)$ & $\hat{\beta_2}(\cdot)$ & $\hat{\beta_3}(\cdot)$ & $\hat{\beta_4}(\cdot)$ & $\hat{\beta_5}(\cdot)$ \\ \hline
% 100         & 43.159                 & 69.041                 & 131.792                & 65.520                 & 31.676                 \\
%             &                        &                        &                        &                        &                        \\
200         & 1.270                  & 2.133                  & 3.830                  & 1.780                  & 1.043                  \\
            &                        &                        &                        &                        &                        \\
400         & 0.218                  & 0.318                  & 0.548                  & 0.284                  & 0.182                  \\
            &                        &                        &                        &                        &                        \\
800         & 0.059                  & 0.081                  & 0.122                  & 0.078                  & 0.054                  \\ \hline
\end{tabular}
\end{table}

\begin{table}[h!]
\caption{Descriptive summaries of the scalar variables considered for the NHANES application}
\label{demo}
\begin{tabular}{lllll}
\hline
\multirow{2}{*}{Characteristic} & \multicolumn{4}{l}{Mean (SD or Proportion)}                                          \\ \cline{2-5} 
                                & Overall (N = 2,816) & Survivor (N = 1,699) & Deceased (N = 1,117) & p-value          \\ \hline
Age                             & 66.0 (9.6)          & 62.1 (8.0)           & 71.8 (8.9)           & \textless{}0.001 \\
Gender                          &                     &                      &                      & \textless{}0.001 \\
\multicolumn{1}{c}{Male}        & 1,431 (51\%)        & 784 (46\%)           & 647 (58\%)           &                  \\
\multicolumn{1}{c}{Female}      & 1,385 (49\%)        & 915 (54\%)           & 470 (42\%)           &                  \\
BMI                             & 28.9 (6.0)          & 29.2 (6.0)           & 28.4 (5.9)           & 0.001            \\
Smoking status                  &                     &                      &                      & \textless{}0.001 \\
\multicolumn{1}{c}{Never}       & 1,247 (44\%)        & 848 (50\%)           & 399 (36\%)           &                  \\
\multicolumn{1}{c}{Former}      & 1,097 (39\%)        & 588 (35\%)           & 509 (46\%)           &                  \\
\multicolumn{1}{c}{Current}     & 472 (17\%)          & 263 (15\%)           & 209 (19\%)           &                  \\ \hline
\end{tabular}
\end{table}

\begin{table}[h!]
\centering
\caption{Selected scalar variables and corresponding hazard ratios}
\begin{tabular}{lllll}
\hline
           & Age   & Gender Female & Former Smoking & Current Smoking \\ \hline
Estimation & 1.096 & 0.749       & 1.313            & 2.171           \\ \hline
           &       &              &                  &                
\end{tabular}
\label{ts}
\end{table}

\section{Web Figures}
\newcommand{\solidline}{\raisebox{2pt}{\tikz{\draw[-,black,solid,line width = 0.3pt](0,0) -- (5mm,0);}}}
\begin{figure}[h!]
    \centering
    \includegraphics[scale=0.8]{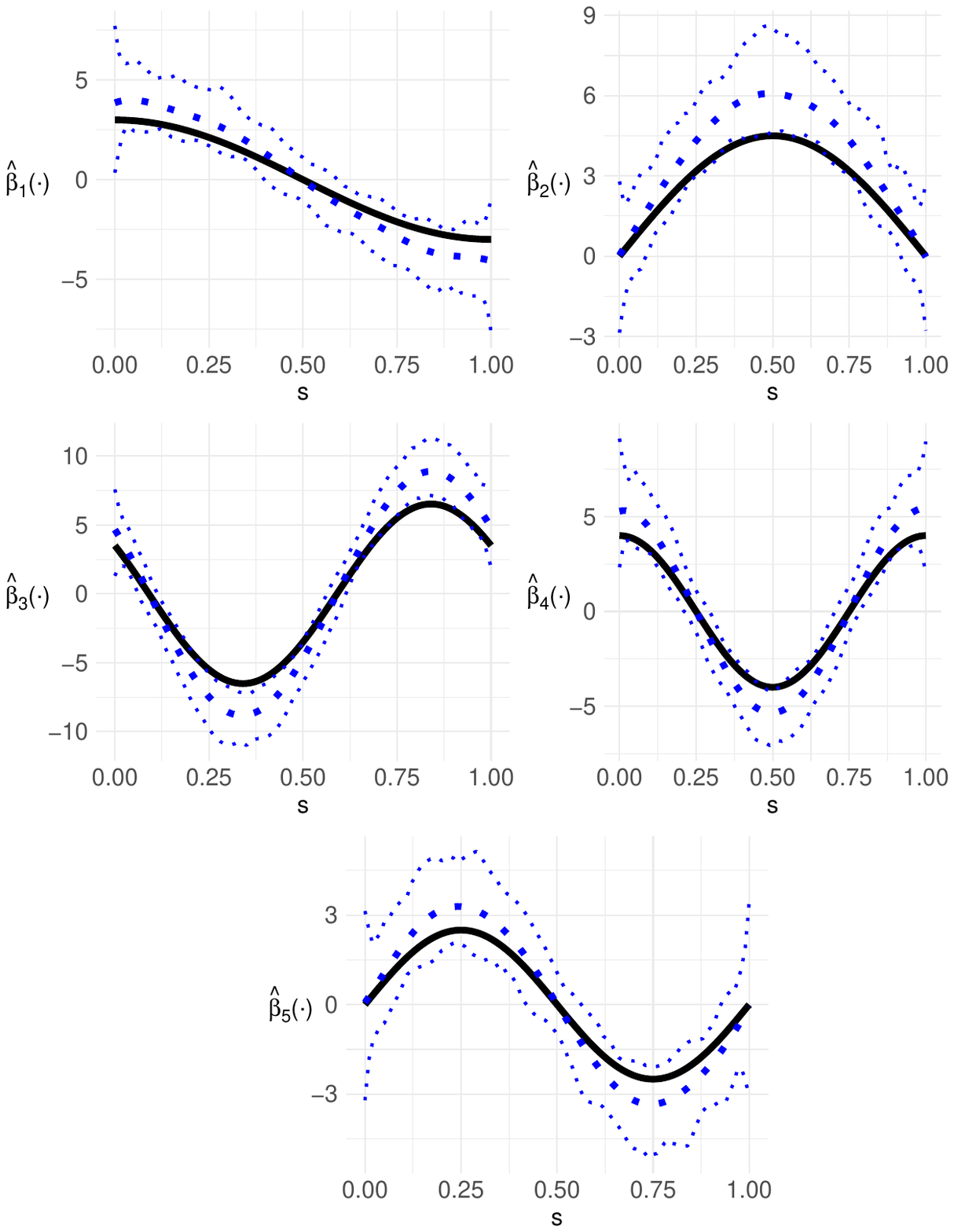}
    \caption{MC estimates and pointwise confidence intervals of the coefficient functions ($n=200$); (\textcolor{blue}{\textbf{$\cdots$}}, VSFCOX; \protect\solidline, true curve)}
    \label{figs200}
\end{figure}

\begin{figure}[h!]
    \centering
    \includegraphics[scale=0.8]{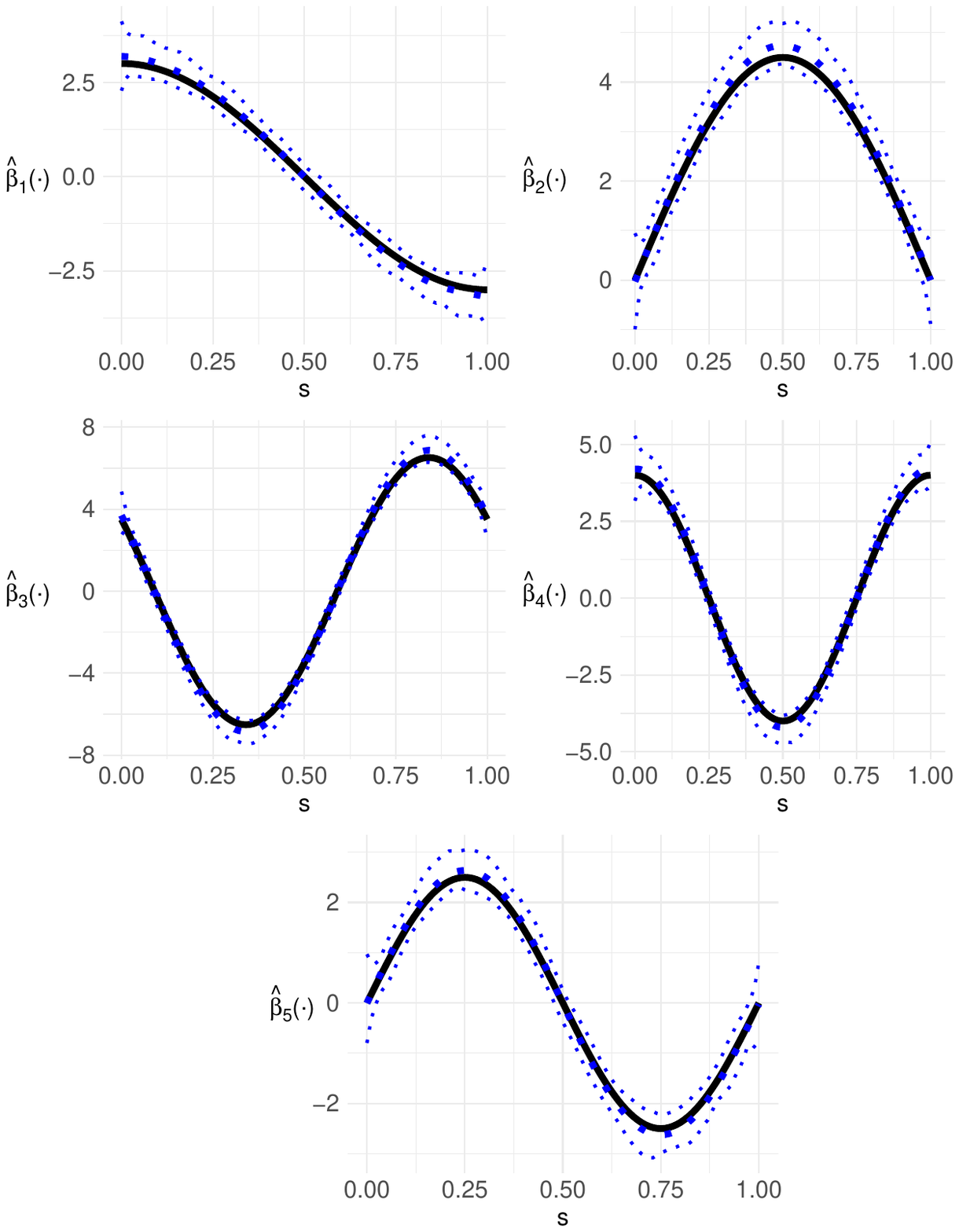}
    \caption{MC estimates and pointwise confidence intervals of the coefficient functions ($n=800$); (\textcolor{blue}{\textbf{$\cdots$}}, VSFCOX; \protect\solidline, true curve)}
    \label{figs800}
\end{figure}

\begin{figure}[h!]
    \centering
    \includegraphics[scale=0.8]{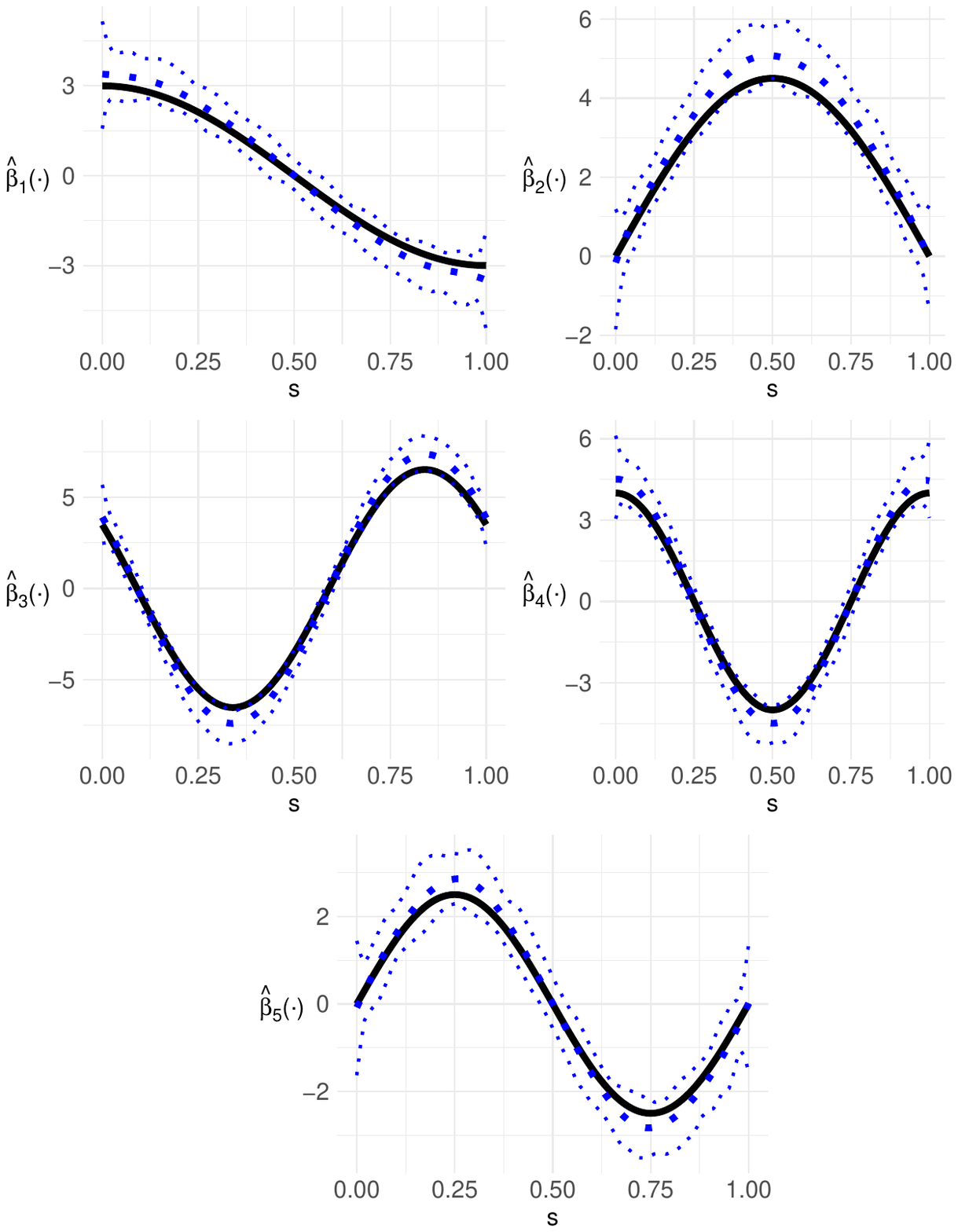}
    \caption{MC estimates and pointwise confidence intervals of the coefficient functions ($n=400$); (\textcolor{blue}{\textbf{$\cdots$}}, VSFCOX; \protect\solidline, true curve) - Adaptive Penalty}
    \label{figsad}
\end{figure}

\begin{figure}[h!]
    \centering
    \includegraphics[scale=0.45]{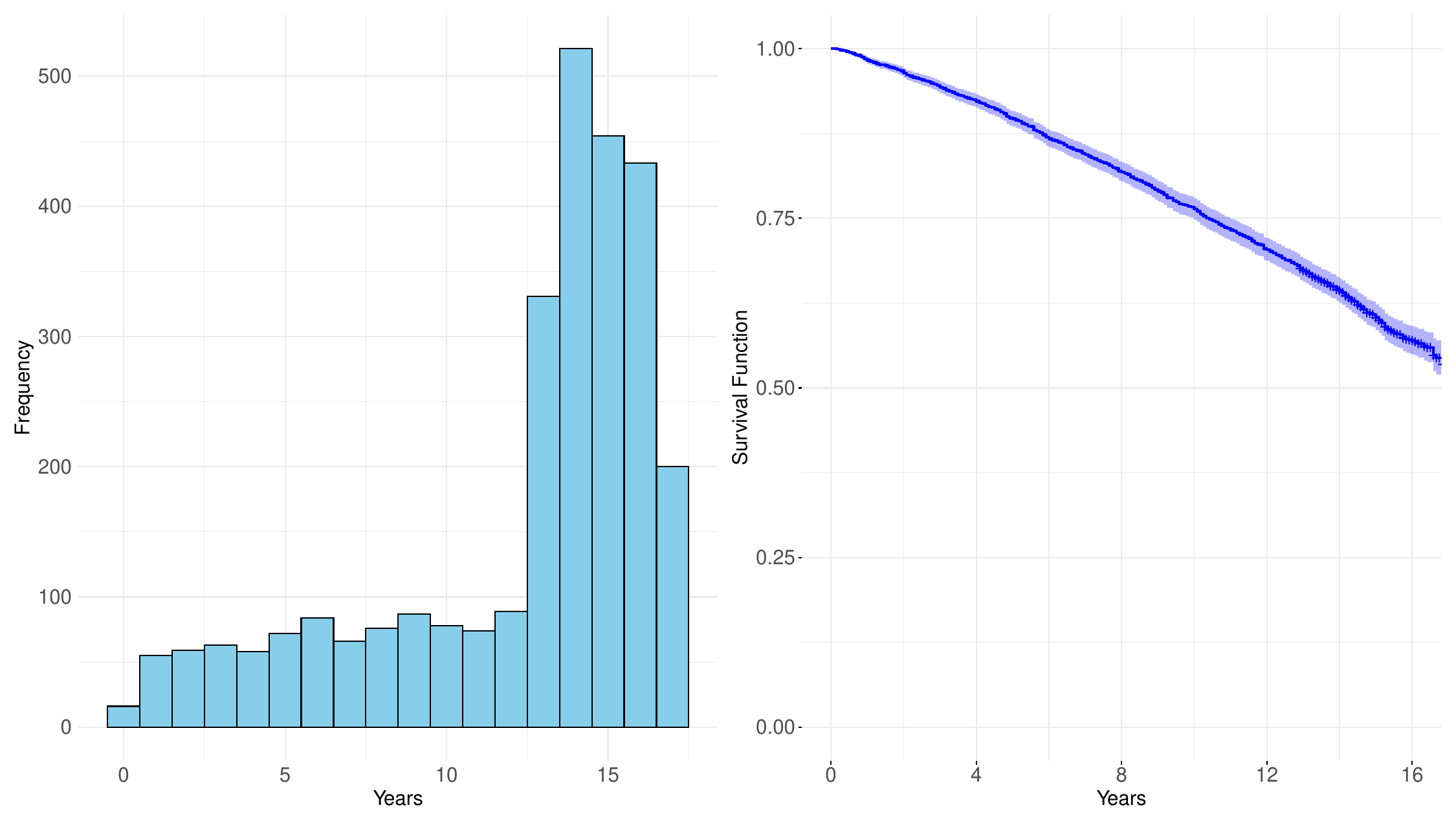}
    \caption{Histogram of survival time and Kaplan-Meier marginal survival curve}
    \label{fig4}
\end{figure}

% \begin{figure}[h!]
%     \centering
%     \begin{subfigure}{0.49\textwidth}
%         \centering
%         \includegraphics[width=\textwidth]{bt1_ad.pdf}
%         \caption{}
%         \label{fig:plot1ad}
%     \end{subfigure}
%     \hfill
%     \begin{subfigure}{0.49\textwidth}
%         \centering
%         \includegraphics[width=\textwidth]{bt2_ad.pdf}
%         \caption{}
%         \label{fig:plot2ad}
%     \end{subfigure}
%     \vspace{0.5cm} % Adjust spacing between rows
%     \begin{subfigure}{0.49\textwidth}
%         \centering
%         \includegraphics[width=\textwidth]{bt3_ad.pdf}
%         \caption{}
%         \label{fig:plot3ad}
%     \end{subfigure}
%     \hfill
%     \begin{subfigure}{0.49\textwidth}
%         \centering
%         \includegraphics[width=\textwidth]{bt4_ad.pdf}
%         \caption{}
%         \label{fig:plot4ad}
%     \end{subfigure}
%     \vspace{0.5cm} % Adjust spacing between rows
%     \begin{subfigure}{0.49\textwidth}
%         \centering
%         \includegraphics[width=\textwidth]{bt5_ad.pdf}
%         \caption{}
%         \label{fig:plot5ad}
%     \end{subfigure}

%     \caption{MC estimates and pointwise confidence intervals of the coefficient functions ($n=400$); (\textcolor{blue}{\textbf{$\cdots$}}, VSFCOX; \protect\solidline, true curve);(a) $\beta_1(\cdot)$; (b) $\beta_2(\cdot)$; (c) $\beta_3(\cdot)$; (d) $\beta_4(\cdot)$; (e) $\beta_5(\cdot)$}
%     \label{figsad}
% \end{figure}

% \newpage
% \bibliographystyle{biom}
% \bibliography{score}